\documentclass[12pt]{iopart}
\usepackage{iopams}
\usepackage{graphicx}
\usepackage{slashed}
\begin{document}
\title{Even-odd parity effects in Majorana junctions}
\author{Alex Zazunov$^1$, Pasquale Sodano$^{2,3}$ and Reinhold Egger$^1$}
\address{$^1$~Institut f\"ur Theoretische Physik, Heinrich-Heine-Universit\"at,
D-40225 D\"usseldorf, Germany}
\address{$^2$~International Institute of Physics, Universidade Federal
do Rio Grande do Norte, 59012-970 Natal, Brazil}
\address{$^3$~INFN, Sezione di Perugia, Via A. Pascoli,
 I-06123 Perugia, Italy}
\ead{egger@thphy.uni-duesseldorf.de}
\begin{abstract}
We study a general Majorana junction, where $N$ helical nanowires
are connected to a common $s$-wave superconductor proximity-inducing Majorana
bound states in the wires.  The normal part of each wire ($j=1,\ldots,N$)
 acts as connected lead, where electrons can tunnel into the respective Majorana
state $\gamma_{A,j}$. The Majorana states at the other end, $\gamma_{B,j}$,
are coupled to each other by an arbitrary tunnel matrix.  We examine the
conditions for even-odd parity effects in the tunnel
conductance for various junction topologies.
\end{abstract}

\submitto{\NJP}

\section{Introduction}

Since their discovery a few years ago, many fascinating
phenomena have been uncovered in topological insulators
\cite{hasan} and topological superconductors \cite{qizhang}.
A particularly simple yet nontrivial example is given by the
Majorana fermion bound states that must be present near the ends
of one-dimensional (1D) topological superconductor wires; for reviews, see
Refs.~\cite{beenakker,alicea,karsten}. Since Majorana fermions
exhibit non-Abelian statistics, they are under
consideration as platforms for topological quantum information
processing.  An experimental realization in terms of few-channel InSb or InAs
nanowires was recently proposed \cite{lutchyn,oreg},
where the conspiracy of pronounced spin-orbit coupling,
strong Zeeman magnetic field, and proximity-induced pairing correlations
inherited from an $s$-wave superconducting substrate, will induce the
Majorana end states.

Electron tunneling from a normal
metal into such a 1D topological superconductor is predicted to
show a zero-bias anomaly conductance peak
due to resonant Andreev reflection
\cite{demler,sodano,nilsson,law,flensberg,wimmer,chung,golub,bose}.
In essence, the Majorana state delocalizes into the normal
metal and forms a resonance pinned to the Fermi energy.
This mechanism establishes a perfect Andreev reflection
channel, with conductance peak value $G=2e^2/h$ at temperature $T=0$.
Resonant Andreev reflection is in fact the most important coupling
mechanism in such a metal-Majorana contact as long as
the proximity-induced gap exceeds the
thermal energy, $k_B T$, and the applied bias voltage scale, $eV$.
All other local perturbations at the contact, e.g., quasi-particle tunneling
or potential scattering, were shown to be irrelevant compared to
tunneling via the resonant Majorana state \cite{fidkowski}.
Experimental reports of conductance peaks with such properties
in InAs or InSb nanowires (with proximity-induced pairing, strong spin-orbit
coupling and appropriate Zeeman fields) have indeed appeared recently
\cite{mourik,exp1,exp2,exp3}, possibly providing first experimental
signatures for Majoranas.  Other Majorana platforms
have also been discussed in the literature; for reviews,
see Refs.~\cite{beenakker,alicea,karsten}.

\begin{figure}[t]
\begin{center}
\includegraphics[width=0.9\textwidth]{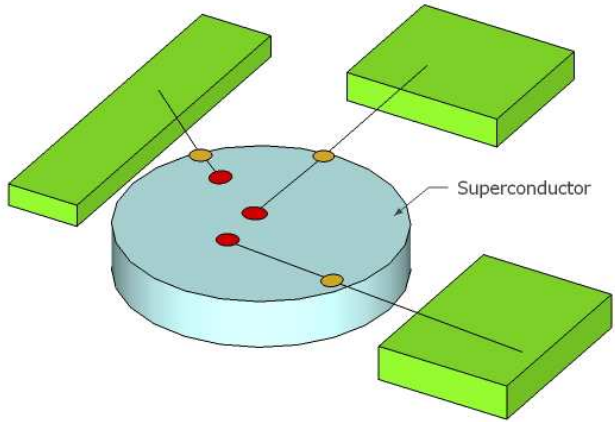}
\end{center}
\caption{\label{fig1}  Schematic Majorana junction of
$N$ helical nanowires (black lines)
connected to a superconducting electrode (blue) and
 $N$ metal electrodes (green); here $N=3$.
The Majorana states $\gamma_{A,j}$ and $\gamma_{B,j}$
are indicated by yellow and red dots,
respectively.  Tunneling between different $\gamma_{B,j}$
is encoded in the antisymmetric matrix $\hat\Omega$, while
$t_j$ refers to the coupling between
$\gamma_{A,j}$ and $\gamma_{B,j}$. Tunneling from the $j$th lead
to $\gamma_{A,j}$ is governed by the hybridization $\Gamma_j\propto
\lambda_j^2$. }
\end{figure}

In order to usefully employ the nonlocal information stored in these
Majorana states, one needs to have junctions where
at least three Majoranas are coupled \cite{alicea1,halperin}.
We note in passing that regular 2D Majorana networks, e.g.,
with many superconducting grains connected by nanowires, can realize
exotic spin models such as Kitaev's toric code \cite{terhal,nussinov}.
In Majorana junctions, the coupling between Majoranas can either be due to
direct tunnel matrix elements, which requires that nanowires are in
close proximity to each other, or mediated through the capacitive Coulomb
charging energy \cite{fu,xu,prb11,heck2,prl12,heck,beri1,hassler,beri2,alex}.
In this paper, we study the Majorana junction shown schematically
in Fig.~\ref{fig1}, where $N$ helical nanowires --- where in the 
simplest case, right-movers have (say) spin up and left-movers spin down 
\cite{hasan} ---  meet
on a superconducting electrode.  Assuming that the superconductor
is grounded, the coupling between Majorana states
is modelled phenomenologically by tunnel matrix elements.
Note that electron-electron interactions in the
normal wire parts may also imply helical Luttinger liquid behavior
\cite{hasan}.
Such effects can be included along the lines of Refs.~\cite{fidkowski,alex},
but here we focus on the simplest noninteracting case.

The setup in Fig.~\ref{fig1} provokes the question whether one can observe
even-odd parity effects in the tunneling conductance.
For tunneling into a chain of Majorana states \cite{flensberg,shivamoggi},
there indeed is a \textit{strong parity effect}: The $T=0$ conductance
exhibits the unitary zero-bias peak value $G_1(0)=2e^2/h$ for tunneling
into a chain with odd number $n$ of coupled Majoranas,
while $G_1(0)=0$ for even $n$.
Moreover, $n-1$ zeros and $n$ peaks are present in the bias-dependent $T=0$
tunneling conductance $G_1(V)$ \cite{flensberg}.
The strong parity effect is explained by noting that $n$ coupled
Majoranas, for even $n$, correspond to $n/2$ finite-energy
fermion states. The absence of a zero-energy state then
implies a quenching of resonant Andreev reflection, and hence $G_1(0)=0$.
For odd $n$, however, one zero-energy Majorana state remains and
allows for  perfect Andreev reflection with $G_1(0)=2e^2/h$.
For suitable choices of the tunnel couplings in Eq.~(\ref{ham0}) below,
we will recover these results for the chain topology in Sec.~\ref{sec3a}.
At the same time, we will extend them by considering other
topologies of the Majorana junction, see Secs.~\ref{sec3b} and
\ref{sec4}.  This allows us to study even-odd parity effects in
a systematic manner within the general setup of Fig.~\ref{fig1}.

The structure of the remainder of this paper is as follows.
In Sec.~\ref{sec2}, we describe the low-energy model of our
Majorana junction and give the formally exact solution for
the steady-state currents flowing through the nanowires.
This solution is elaborated for the ``chain'' and ``loop'' topologies
of the junction in Sec.~\ref{sec3}, based on analytical
results for $N=2$ and $N=3$, combined with numerics for arbitrary $N$.
In Sec.~\ref{sec4}, we show that analytical progress is possible
for additional junction topologies, where $N$ can be arbitrary but
$\Gamma_j=\Gamma$ and $t_j=t$.  In particular, we address the
possibility of parity effects in the tunneling conductance for the ``ladder'',
``tree,'' and ``isotropic'' junction topologies.
Finally, we offer a general interpretation of our results in Sec.~\ref{sec5}.
Some details have been delegated to the Appendix.

\section{Multi-terminal Majorana Junction}\label{sec2}

\subsection{Model}

We consider a junction of $N$ helical nanowires ($j=1,\ldots,N$) assembled
on top of a conventional $s$-wave superconducting electrode,
see Fig.~\ref{fig1} for a schematic illustration of this multi-terminal
Majorana junction.  We assume that the superconductor is grounded, and hence
charging effects are irrelevant.  By virtue of the proximity
effect, each nanowire segment located on top of the superconductor
turns into a 1D topological superconductor \cite{alicea}.
The bulk-boundary correspondence then implies that there must be
a single unpaired Majorana fermion, $\gamma_{B,j}$, at
the end of the superconducting wire,
plus another one, $\gamma_{A,j}$,
at the interface to the normal part of the wire. The latter part also acts
as connected lead and allows to probe electron transport through the
multi-terminal Majorana junction in Fig.~\ref{fig1}.
The Majoranas correspond to self-conjugate operators,
$\gamma_{\alpha,j}^\dagger=\gamma_{\alpha,j}^{}$  (with $\alpha=A,B$),
subject to the anticommutation relations
\begin{equation}
\{ \gamma_{\alpha,j} , \gamma_{\alpha',j'} \} = \delta_{\alpha\alpha'}
\delta_{jj'}.
\end{equation}
It is convenient to combine them to $N$-component objects,
$\gamma_{\alpha}=(\gamma_{\alpha,1},
\ldots,\gamma_{\alpha,N})^T$. Similarly, the conventional lead fermion
operators are contained in $c=(c_1,\ldots,c_N)^T$, with
$c_j = \sum_k c_{jk}$, where $c_{jk}$ annihilates a lead fermion of
momentum $k$ in the $j$th lead. For simplicity, we assume identical
dispersion relation $\xi_k$ for all leads, and
denote their respective chemical potentials as $\mu_j$ (with
$\mu_S=0$ for the superconductor).

Throughout this paper, we focus on the most interesting case where
the proximity-induced pairing gap exceeds both the thermal energy, $k_B T$,
and the applied bias energies, such that quasi-particle excitations above
the gap are negligible.  Retaining only the Majoranas as relevant
fermionic degrees of freedom
for the junction in Fig.~\ref{fig1}, the Hamiltonian reads
\begin{equation}\label{ham0}
H = H_l+ \frac{i}{2} \gamma_B \hat \Omega \gamma_B + i\gamma_A \hat t \gamma_B
+ ( c^\dagger -c ) \hat \lambda \gamma_A ,
\end{equation}
where $H_l=\sum_{j=1}^N \sum_{k} \xi_k c^\dagger_{jk} c_{jk}^{}$
describes the leads.
The ``hat'' notation in Eq.~(\ref{ham0}) refers to an $N\times N$
matrix structure in wire-number $(j)$ space.
Without loss of generality, the antisymmetric
matrix $\hat\Omega$, containing the couplings between
different $\gamma_{B,j}$, is chosen real-valued.
The same holds true for the tunnel couplings $t_j$,
connecting $\gamma_{A,j}$ and $\gamma_{B,j}$ in the $j$th wire,
and for $\lambda_j$, connecting $\gamma_{A,j}$ to the
lead fermion $c_j$;  we use
$\hat t={\rm diag}(t_1,\ldots,t_N)$ and
$\hat \lambda=(\lambda_1,\ldots, \lambda_N)$.  Within the standard
wide-band approximation for the leads \cite{alexa}, the $\lambda_j$ only appear
through the hybridization scales $\Gamma_j= 2\pi \nu_0 \lambda_j^2,$
where $\hat\Gamma=(\Gamma_1,\ldots,\Gamma_N)$ and
$\nu_0=\sum_k \delta(\xi_k)$ is the constant lead density of states.
By putting some of the couplings $\{ \Gamma_j,t_j,\Omega_{j<j'}
\}$ to zero, Eq.~(\ref{ham0}) serves as a general model for $M\le N$
normal-metal terminals connected through a junction of
$n\le 2N$ coupled Majorana fermions.
The aim of this paper is to quantitatively understand the tunneling
current into such a multi-terminal Majorana junction.

\subsection{Conductance}

The stationary current flowing in the $j$th nanowire towards the
junction, $I_j$, follows from the relation \cite{flensberg,prl12}
\begin{equation}\label{meirwin}
I_j = \frac{e\Gamma_j}{h} \int d\epsilon \  F(\epsilon - \mu_j)
\ {\rm Im}  {\cal G}^r_{Aj,Aj}(\epsilon),
\end{equation}
where we have Fermi distributions in the leads,
$F(\epsilon) =\tanh(\epsilon/2 k_B T)$, and
${\cal G}^r_{\alpha j,\alpha'j'}(\epsilon)$ is
the energy representation of the retarded Majorana Green's function.
The main quantity of interest in experiments is the corresponding
differential conductance $G_j = e \partial I_j/\partial \mu_j$,
which yields from Eq.~(\ref{meirwin}) the $T=0$ result
\begin{equation}\label{diffcondzero}
G_j = \frac{2e^2}{h} \Gamma_j [ - {\rm Im}{\cal G}^r_{Aj,Aj}(\mu_j) ].
\end{equation}
Since we have a noninteracting Hamiltonian in Eq.~(\ref{ham0}),
the Dyson equation for ${\cal G}^r$ can be solved exactly.
As sketched in the Appendix, the $2N\times 2N$
matrix for ${\cal G}^r$ thereby takes the form
(the $2\times 2$ block structure refers to $\alpha=A,B$ space)
\begin{equation}\label{gret}
{\cal G}^r (\epsilon) =  \left(
\begin{array}{cc} \epsilon\hat 1+i\hat \Gamma & -i\hat t\\
i\hat t & \epsilon \hat 1-i\hat \Omega\end{array} \right)^{-1},
\end{equation}
where $\hat 1$ is the $N\times N$ unit matrix in wire-number space.
The matrix element ${\cal G}^r_{Aj,Aj}(\epsilon)$ obtained from
Eq.~(\ref{gret}) then determines the current $I_j$
and hence the conductance $G_j$.

Current conservation, $\sum_{j=1}^N I_j=0$, is \textit{not}\
imposed here due to the presence of a grounded superconductor \cite{demler}.
Since the retarded Green's function in Eq.~(\ref{gret}) does not depend
on the chemical potentials, the current $I_j$, and hence
also the conductance $G_j$, depends on the respective chemical
potential $\mu_j$ only, but not on the $\mu_{k\ne j}$.  All ``nonlocal''
conductances, $e\partial I_j/\partial \mu_{k\ne j}$,
therefore vanish identically regardless of the
values of the tunnel couplings,  as long as the validity requirements for
Eq.~(\ref{ham0}) are met.
The resulting decoupling of different terminals is a characteristic
feature of noninteracting Majorana networks and can be traced back to
the assumption of a {\it grounded}\ superconductor.
For a ``floating'' (not grounded) mesoscopic superconductor,
where current must be conserved and Coulomb charging effects may become important,
nonlocal conductances are typically finite.

\begin{figure}[t]
\begin{center}
\includegraphics[width=0.8\textwidth]{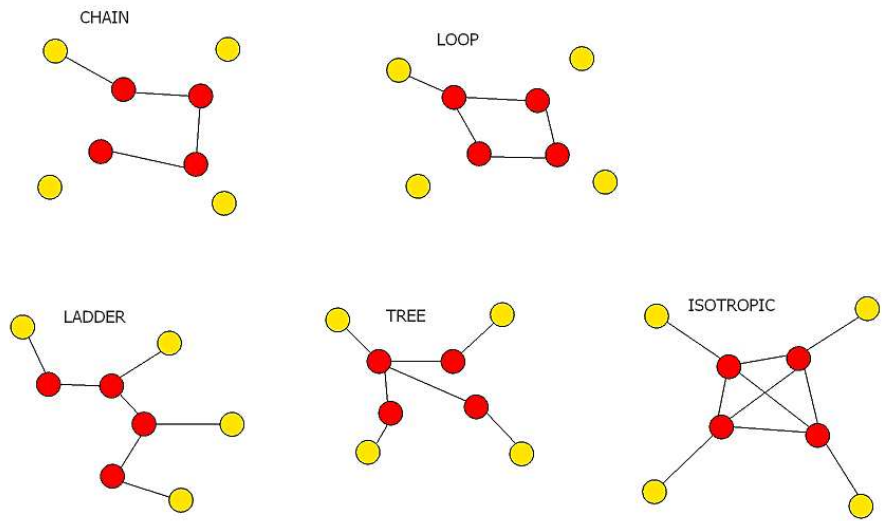}
\end{center}
\caption{\label{fig2}  Different topologies of a Majorana junction,
shown here for $N=4$.
Yellow (red) dots refer to $\gamma_{A,j}$ ($\gamma_{B,j})$
Majorana states, with the respective top left state for $j=1$.
A line connecting two dots indicates that the respective coupling
$t_j$ or $\Omega_{j,k}$ is finite.  }
\end{figure}

From now on, without loss of generality,  we consider only the tunneling
conductance $G_1(V)=dI_1(V)/dV$ for nanowire $j=1$, with $eV\equiv \mu_1$
and in the most interesting $T=0$ limit,
 see Eq.~(\ref{diffcondzero}).
It is worth stressing that while $G_1$ is independent of
the chemical potentials $\mu_{j>1}$, it still does depend on all hybridization
parameters $\Gamma_{j}$.  Although $G_1(V)$ can always be expressed in terms
of the retarded Green's function (\ref{gret}),  explicit results for
arbitrary parameters are generally lengthy and not illuminating.
We here instead aim at analytical results by
examining several limiting cases for the specific junction topologies
drawn in Fig.~\ref{fig2}.  In Sec.~\ref{sec3},
we consider $N=2$ nanowires with arbitrary coupling matrix
elements, where a compact expression for $G_1(V)$ allows us to study
tunneling into a ``chain'' of Majoranas.
The case of $N=3$ wires is also addressed, where we obtain insights
for the ``loop'' geometry in Fig.~\ref{fig2}.
We then analyze the linear conductance $G_1(0)$
for loop configurations with arbitrary $N>2$ Majorana states in the loop
and study the parity effect for such a junction topology.
In Sec.~\ref{sec4}, we discuss the remaining three
topologies shown in Fig.~\ref{fig2}, where we assume
$\Gamma_j=\Gamma$ and $t_j=t$ in order to simplify the analysis.
In addition, all non-zero matrix entries in $\hat\Omega$
are assumed equal, i.e., $\Omega_{j<k}$ is either some finite constant
($\Omega$) or zero. Under these conditions, we show that analytical results
for $G_1(V)$ with arbitrary $N$ can be derived,
which allows us to systematically study even-odd parity effects in
Majorana junctions.  Before moving on, however, we
first discuss a useful reformulation of Eq.~(\ref{gret}).

\subsection{Retarded Green's function}

To achieve analytical progress for arbitrary $N$, at least
for some special parameter choices in Eq.~(\ref{ham0}), it is convenient to
write the relevant matrix element, ${\cal G}_{A1,A1}^r(\epsilon)$,
of the retarded Green's function in an alternative form
avoiding direct matrix inversion.
We first single out the effects of the inter-wire couplings $\hat \Omega$
by writing a Dyson equation for ${\cal G}^r$,
\begin{equation}\label{newdyson}
{\cal G}^r = {\cal G}_0^r+ {\cal G}_0^r \left(
\begin{array}{cc} 0 & 0 \\ 0 & i\hat \Omega \end{array}\right)
{\cal G}^r,
\end{equation}
where the ``unperturbed'' ($\hat \Omega=0$)
retarded Green's function ${\cal G}_0^r$ is diagonal in wire-number space.
In fact, we can read off its matrix elements from Eq.~(\ref{gret}),
\begin{equation}\label{g0ret}
[{\cal G}^r_0(\epsilon)]_{\alpha j, \alpha' j'} = \frac{\delta_{jj'}}
{ \epsilon^2-t_j^2+i\epsilon\Gamma_j}
\left( \begin{array}{cc}
\epsilon & i t_j \\ -i t_j & \epsilon+i\Gamma_j
\end{array}\right)_{\alpha,\alpha'}.
\end{equation}
Now we introduce the auxiliary quantities
\begin{equation}
X_{j}(\epsilon) \equiv \frac{{\cal G}_{Bj,A1}^r(\epsilon) }
{[{\cal G}_0^r]_{B1,A1}(\epsilon) } ,
\end{equation}
which encode the off-diagonal matrix elements of the retarded
Green's function and fulfill the relation
\begin{equation}\label{xx}
X_j -i \frac{\epsilon+i\Gamma_j}{\epsilon^2-t_j^2+i\epsilon\Gamma_j}
\sum_{k=1}^N \Omega_{jk} X_k = \delta_{j,1}.
\end{equation}
Equation (\ref{xx}) follows by taking the $(Bj,A1)$ matrix element
of Eq.~(\ref{newdyson}), and
is widely used below to determine the $X_j$.
Note that for the tunneling conductance (\ref{diffcondzero})
we need ${\cal G}_{A1,A1}^r(\epsilon)$.
Taking the $(A1,A1)$ matrix element of Eq.~(\ref{newdyson}) yields after
some algebra
\begin{equation}\label{ga1a1}
{\cal G}^r_{A1,A1}(\epsilon) = \frac{1}{\epsilon+i\Gamma_1}
\left( 1 + \frac{t_1^2 X_1 (\epsilon) }{\epsilon^2-t_1^2+
i\epsilon\Gamma_1} \right).
\end{equation}
Equation (\ref{ga1a1}) can be further simplified for
$\epsilon=eV\to 0$, where the \textit{linear conductance} is
\begin{equation}\label{lincond}
G_1(0) = \frac{2e^2}{h} ( 1-X_1).
\end{equation}
The real-valued $X_1=X_1(\epsilon=0)$ follows from solving Eq.~(\ref{xx}).

\section{Chain and loop junctions}\label{sec3}

\subsection{Conductance for chain topology} \label{sec3a}

Let us start with the case of $N=2$ nanowires,
where $\hat \Omega$ is fully
determined by just one parameter, $\Omega_{12}=\Omega$.
Equation (\ref{gret}) then yields the
Green's function matrix element determining the
tunneling conductance,
\begin{equation}\label{gret2}
{\cal G}_{A1,A1}^r(\epsilon)=
\frac{\epsilon t_2^2 + (\Omega^2-\epsilon^2)(\epsilon+i\Gamma_2)}{
\Omega^2(\epsilon+i\Gamma_1)(\epsilon+i\Gamma_2)-
(\epsilon^2-t_1^2+i\epsilon\Gamma_1)(\epsilon^2-t_2^2+i\epsilon\Gamma_2) }.
\end{equation}
Several remarks about this result are in order.
(i) For $t_1=0$, Eq.~(\ref{gret2}) reduces
to ${\cal G}^r_{A1,A1}=1/(\epsilon+i\Gamma_1)$, which leads to the
well-known resonant Andreev reflection peak for tunneling into a single
decoupled Majorana fermion ($\gamma_{A1}$) \cite{demler,law},
\begin{equation}\label{rar}
G_1 (V)= \frac{2e^2}{h} \frac{\Gamma_1^2}{\Gamma_1^2+(eV)^2}.
\end{equation}
Note that the parameters $t_2$ and $\Gamma_2$ do not affect this result at all.
(ii) Without coupling between the wires, $\Omega=0$,
but with $t_1\ne 0$, the Majorana $\gamma_{B,1}$ is also
involved and the resonance (\ref{rar}) will split into two symmetric
finite-voltage peaks separated by a perfect dip at zero bias
 \cite{demler,law}. Indeed, now Eq.~(\ref{gret2})  yields
\begin{equation}\label{gome}
G_1 (V) = \frac{2e^2}{h}
\frac{(eV\Gamma_1)^2}{(eV\Gamma_1)^2+ [(eV)^2-t_1^2]^2},
\end{equation}
such that the linear ($V\to 0$) conductance vanishes but
two perfect ($G_1=2e^2/h$) peaks are present for $eV=\pm t_1$.
This reflects the parity effect in a chain of $n$ coupled
Majoranas, where $G_1(V)$ has $n-1$ zeroes
and $n$ unitary peaks \cite{flensberg},  in accordance with
 Eq.~(\ref{gome}) for $n=2$.
(iii) Putting $t_2=0$ in Eq.~(\ref{gret2}), we effectively recover
tunneling into the $n=3$ Majorana chain
composed by $\gamma_{A1}-\gamma_{B1}- \gamma_{B2}$,
\begin{equation}
G_1 (V) = \frac{2e^2}{h} \frac{\Gamma_1^2[\Omega^2- (eV)^2]^2}{
(eV)^2 [\Omega^2+t_1^2-(eV)^2]^2+\Gamma_1^2 [\Omega^2-(eV)^2]^2},
\end{equation}
which exhibits three unitary conductance peaks, at
$V=0$ and $eV=\pm\sqrt{\Omega^2+t_1^2}$, and two
zeroes, $eV=\pm \Omega$,  as expected under the parity effect.
(iv) Finally, the $n=4$ Majorana chain is realized
for $\Gamma_2=0$ in Eq.~(\ref{gret2}).
For simplicity, we put $t_1=t_2=t$ and obtain
\begin{equation}
G_1 = \frac{2e^2}{h} \frac{( eV\Gamma_1)^2[\Omega^2+t^2- (eV)^2]^2}{
\left[(eV)^2\Omega^2- ((eV)^2-t^2)^2\right]^2
+(eV\Gamma_1)^2  [\Omega+t^2-(eV)^2]^2}.
\end{equation}
Clearly, we have three zeroes, $V=0$ and $eV=\pm \sqrt{\Omega^2+t^2}$,
and four unitary conductance peaks, $eV=\pm\frac{\Omega}{2}\pm
\sqrt{\Omega^2/4+t^2}$ with independent $\pm$ signs.
When also $\Gamma_2\ne 0$, the parity effect is not ideal
anymore, i.e. peak conductances are smaller than $2e^2/h$,
but it remains observable.

The above Majorana chain configuration can be generalized
to setups with $N>2$ and $n\le N+1$ Majorana states, see
Fig.~\ref{fig2}.  Let us focus on computing 
the linear conductance $G_1(0)$ for $n=N+1$.
To that end, we put $t_{j>1}=0$ and keep as non-zero entries in
$\hat\Omega$ only the nearest-neighbor couplings $\Omega_{j,j+1}$
with $j=1,\ldots,N-1$. 
In the limit $\epsilon \to 0$, the Dyson equation (\ref{xx}) yields
\begin{equation}\label{Xchain}
X_1 - \frac{\Gamma_1 \Omega_{12}}{ t_1^2} \ X_2  = 1 , \quad
\Omega_{j,j+1} X_{j+1} + \Omega_{j,j-1} X_{j-1} = 0
\end{equation}
for $j = 2,\ldots, N$.  The boundary condition $X_{N+1} = 0$ leads to
$X_{j\,\, {\rm odd}} = 0$  ($X_{j\,\, {\rm even}} = 0$)
for even (odd) $N$, 
and hence $X_1 = \delta_{N, {\rm odd}}$. 
The conductance (\ref{lincond}) for tunneling into
 this chain of $n=N+1$ Majoranas is
\begin{equation}\label{G1chain}
G_1(0) =\frac{2 e^2}{ h} \, \delta_{n, {\rm odd}} ,
\end{equation}
and exhibits a strong parity effect.
Moreover, we have numerically confirmed 
the strong parity effect in such a junction
by computing the differential conductance.
In full agreement with Ref.~\cite{flensberg},
we found that $G_1(V)$ exhibits $n$ unitary-limit ($2 e^2/h$) peaks
as a function of bias $V$.

\subsection{Conductance for loop topology }\label{sec3b}

\begin{figure}[t]
\begin{center}
\includegraphics[width=0.8\textwidth]{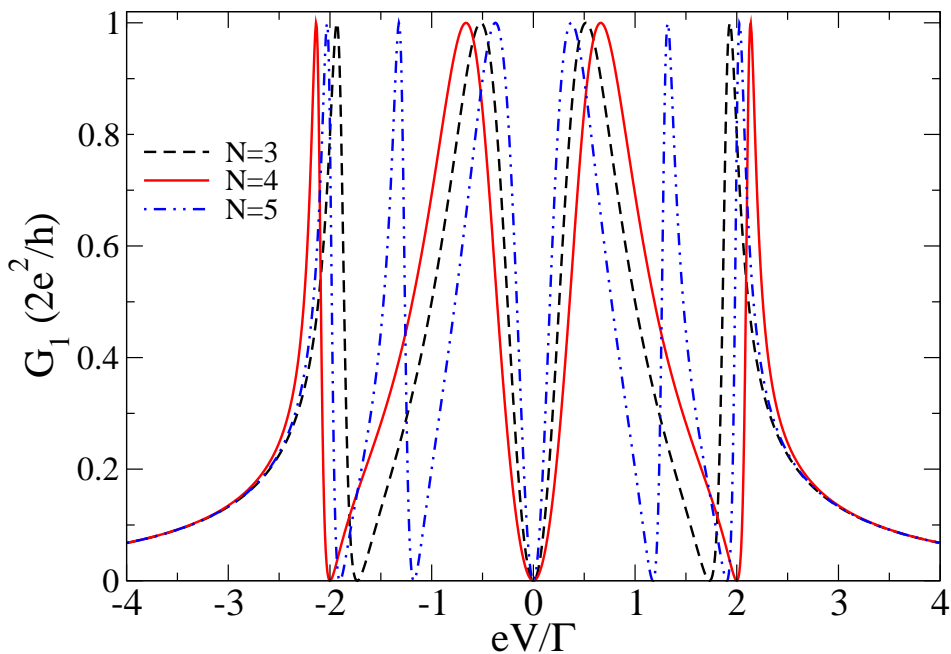}
\end{center}
\caption{\label{fig3}  Tunneling conductance $G_1(V)$ \textit{vs.}\ $eV/\Gamma$
(with $\Gamma=\Gamma_1$)
for a Majorana junction in the homogeneous ``loop'' configuration
($\Omega_{j, j + 1} = \Omega$), shown for several $N$ with
 $t_1/\Gamma= \Omega /\Gamma=1$.  }
\end{figure}

The algebra needed to obtain $G_1(V)$ from Eq.~(\ref{gret})
for $N>2$ Majorana wires becomes much more involved.
For $N=3$, the linear conductance can still be given
in compact form for arbitrary parameters. We find:
\begin{equation}\label{glin3}
G_1(0) = \frac{2e^2}{h} \frac{\Gamma_1 \xi}{\Gamma_1\xi+t_1^2 (
t_2^2 t_3^2+\Gamma_2\Gamma_3\Omega_{23}^2)}, \quad
\xi = \Omega_{12}^2\Gamma_2 t_3^2 + \Omega_{13}^2\Gamma_3 t_2^2.
\end{equation}
For $t_1\ne 0$, the unitary limit  ($G_1=2e^2/h$) is reached when either
$t_2=\Gamma_3\Omega_{23}=0$ or $t_3=\Gamma_2\Omega_{23}=0$, while
at the same time $\xi\ne 0$.
Note that $G_1(0)=0$ when $t_2=t_3=0$.  The junction then consists
of $\gamma_{A1}$ coupled to a closed ``loop'' consisting of the three
$B$ Majoranas.  This is the simplest example for the ``loop''
topology shown in Fig.~\ref{fig2}.

The linear conductance $G_1(0)$ for loop junction configurations
with arbitrary $N > 2$ and non-vanishing $\Omega_{j,j + 1}$
can readily be analyzed using the Dyson equation (\ref{xx}).
As before, see Sec.~\ref{sec3a}, assuming $t_j = t_1 \delta_{j, 1}$ and taking
the limit $\epsilon \to 0$ in Eq.~(\ref{xx}), we obtain
\begin{equation}\label{Xloop}
X_1 - \frac{\Gamma_1}{t_1^2} 
\left( \Omega_{12} X_2 + \Omega_{1 N} X_N \right)  = 1, \quad
\Omega_{j,j+1} X_{j+1} + \Omega_{j,j-1} X_{j-1} = 0
\end{equation}
for $j = 2,\ldots, N$, with periodic boundary conditions, 
$X_{N+1} = X_{1}$.
For odd $N$, one can verify that Eq.~(\ref{Xloop})
has always, i.e., for arbitrary $\Omega_{j, j+1}$, the solution $X_1 = 1$.
Indeed, for $X_1 = 1$, the two sequences $X_{j \,\, {\rm odd / even}}$,
both obeying
\begin{equation}
X_j = \frac{\Omega_{j-2,j-1}}{ \Omega_{j-1,j}} \, X_{j-2}
\end{equation}
with $j \neq 1 \, ({\rm mod}\, N)$, are uniquely determined
by virtue of the matching relation 
$\Omega_{12} X_2 + \Omega_{1 N} X_N = 0$.
As a result, the linear conductance (\ref{lincond}) 
always vanishes for odd $N$.
For even $N$, a general solution to Eq.~(\ref{Xloop}) is given by
\begin{eqnarray}\label{Xjloop}
X_{j\,\, {\rm odd}} &=& 0   ,\\ \nonumber
X_{j\,\, {\rm even}}& =& \frac{\Omega_{j-2,j-1}}{ 
\Omega_{j-1,j}} \, X_{j-2} ,\\ \nonumber
\Omega_{12} X_2 &+& \Omega_{1 N} X_N  = - t_1^2 / \Gamma_1 ,
\end{eqnarray}
implying that the linear conductance (\ref{lincond}) reaches
the unitary-limit value $2 e^2 / h$.
This suggests that like for a Majorana chain,
the strong parity effect described by Eq.~(\ref{G1chain}) with $n=N+1$
is also present for the loop topology.
Thus, by changing the number of tunnel coupled nanowires $N$
in the loop configuration, one may switch the conductance $G_1(0)$ on and off,
depending on the parity of $N$.  
This parity-based switching mechanism for the tunneling conductance
stems from the non-local nature of electron transport in Majorana junctions,
and offers a way to ``engineer'' such parity effects.

\begin{figure}[t]
\begin{center}
\includegraphics[width=0.8\textwidth]{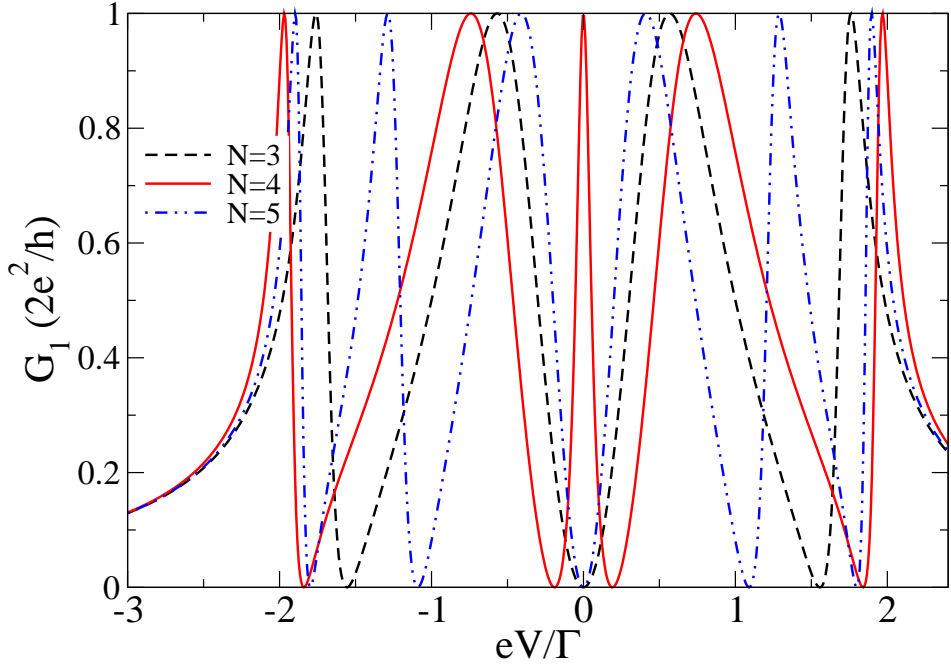}
\end{center}
\caption{\label{fig4} Same as in Fig.~\ref{fig3} but for generic 
inhomogeneous ``loop'' configuration, where only 
$\Omega_{1,2}=0.65\Gamma_1$ has been changed; 
all other $\Omega_{i,i+1}=\Gamma_1$
as in Fig.~\ref{fig3}.  }
\end{figure}

However, the solution (\ref{Xjloop}) fails when $N$ is even and the
parameters $\Omega_{j,j+1}$ satisfy the incompatibility condition
\begin{equation}\label{incompat}
\Omega_{12} + \Omega_{1 N} \prod_{k=1}^{N/2 - 1}
\frac{\Omega_{2k,2k+1}}{\Omega_{2k+1,2k+2}} = 0 .
\end{equation}
In particular, for a homogeneous loop configuration 
with identical couplings [$\Omega_{j, j + 1} = \Omega$
for all $j$], Eq.~(\ref{incompat}) is satisfied.
We then have the trivial solution $X_j = 1$ for all $j$,
 regardless of the parity of $N>2$.
Consequently, $G_1(0)=0$ for arbitrary (odd or even) 
number of Majoranas in such a loop.
This can be understood as a result of the
destructive interference -- with complete cancellation of clockwise and
anti-clockwise contributions to $G_1(0)$ -- in a 
homogeneous Majorana loop (closed chain).

Remarkably, for even $N$ and arbitrary Majorana loops, this interference 
therefore results in either a maximal ($G_1=2e^2/h$) or a 
completely blocked ($G_1=0$) zero-bias conductance,
depending on the incompatibility condition in Eq.~(\ref{incompat}).
Fig. \ref{fig3} shows the differential conductance $G_1(V)$ for several $N$
in the homogeneous loop configuration.  Numerically, we find that
the number $P$ of finite-bias peaks in $G_1(V)$ for this configuration
is given by
\begin{equation}\label{Ploop}
P= \left\{ \begin{array}{rr} N+1, &  N \, {\rm odd},\\
                            2(1+[N/4]),& N \, {\rm even},\end{array}\right.
\end{equation}
with $[x]$ denoting the integer part of $x$.
Due to the lack of an unambiguous correspondence between
$P$ and $N$, a parity effect is difficult to detect in the homogeneous 
loop. 

Let us stress again that in the loop configuration, $G_1(V)$ exhibits 
the strong parity effect,
with $N+1$ unitary-limit ($2 e^2 / h$) peaks as in the chain configuration,
for generic (inhomogeneous) parameters.
This is illustrated in Fig.~\ref{fig4} for a particular choice
of the $\Omega_{i,i+1}$ couplings.

\section{Junction topologies and parity effects}\label{sec4}

In this section, we address the three remaining junction topologies in
Fig.~\ref{fig2} for arbitrary $N$.
The tunneling conductance is obtained for a parameter set with
equal hybridizations, $\Gamma_j=\Gamma$, and intra-wire couplings, $t_j=t$.
The matrix elements $\Omega_{j<k}$ are either zero or equal to
a single constant ($\Omega$), depending on the junction type.

\subsection{Ladder configuration}\label{sec4a}

For the ``ladder'' shown in Fig.~\ref{fig2},
the only non-zero $\Omega_{j<k}$ entries are nearest-neighbor bonds,
$\Omega_{j,j+1}=\Omega$ for $j=1,\ldots,N$ (``closed ladder'')
or $j=1,\ldots, N-1$ (``open ladder'').  The example shown in Fig.~\ref{fig2}
refers to an open ladder for $N=4$.
For both ladder configurations, Eq.~(\ref{xx})
for $X_j(\epsilon)$ takes the form
\begin{equation}\label{xxxx}
X_j- A (X_{j+1}-X_{j-1}) = \delta_{j,1},\quad
A(\epsilon) = \frac{i\Omega(\epsilon+i\Gamma)}{\epsilon^2-t^2+i\epsilon\Gamma},
\end{equation}
supplemented by the boundary conditions $X_0=X_{N+1}=0$
($X_j=X_{j+N})$ for the open (closed) ladder configuration.

\begin{figure}[t]
\begin{center}
\includegraphics[width=0.9\textwidth]{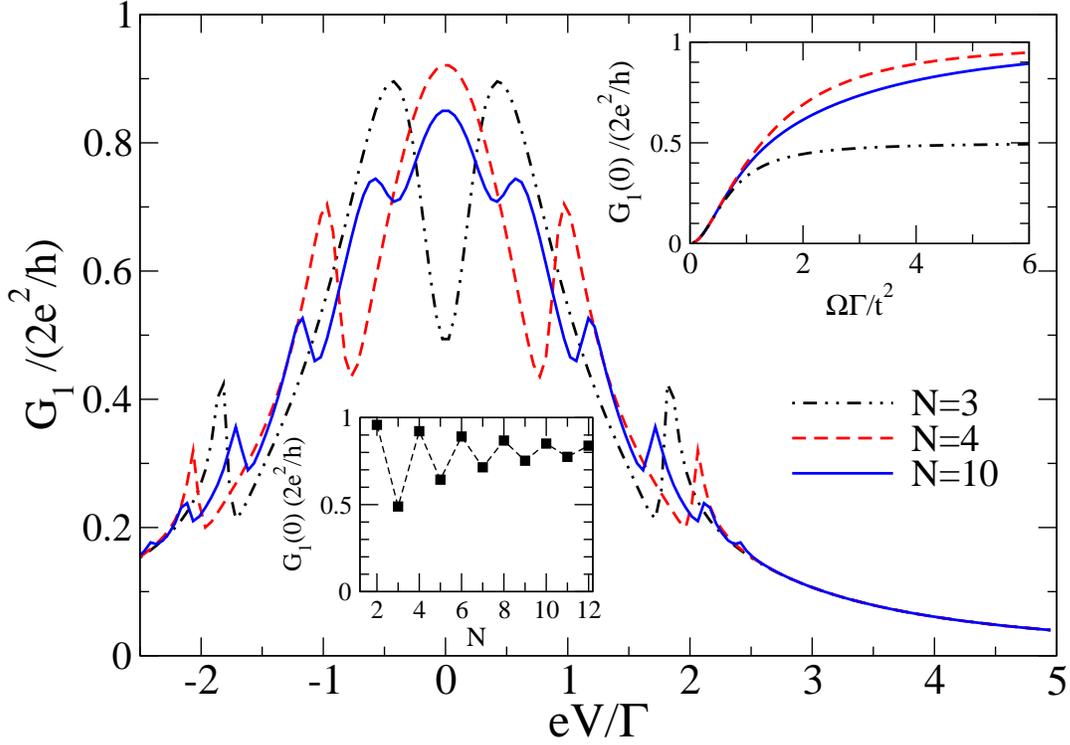}
\end{center}
\caption{\label{fig5}  Differential conductance $G_1(V)$ for tunneling
into a Majorana junction in the ``open ladder'' configuration.
Main panel: $G_1$ \textit{vs.}\ $eV/\Gamma$ with
$t/\Gamma=0.5, \Omega/\Gamma=1.2$ and several $N$.
Upper right inset: Linear conductance $G_1(0)$
\textit{vs.}\ $\Omega \Gamma/t^2$ for $N=3,4,$ and $10$.
Lower inset: $N$-dependence of $G_1(0)$ for same parameters.}
\end{figure}

Starting with the \textit{open ladder}, the solution of Eq.~(\ref{xxxx})
is
\begin{equation}\label{xopen}
X_1(\epsilon) = \frac{\sqrt{1+4A^2}}{2A^2} \frac{F_+^{N+1}
+F_-^{N+1}}{F_+^{N+1}-F_-^{N+1}} - \frac{1}{2A^2} ,
\quad F_\pm(\epsilon) = \frac{1\pm \sqrt{1+4A^2}}{2A}.
\end{equation}
The conductance $G_1(V)$ follows readily from Eqs.~(\ref{diffcondzero})
and (\ref{ga1a1}).  For $N\gg 1$, using $|F_+/F_-|>1$, we find that
$X_1$ in Eq.~(\ref{xopen}) becomes $N$-independent,
$X_1\simeq -F_-/A.$  Hence $N$-dependent
parity effects must disappear in this junction type for
large $N$.  However, for moderate $N$, we still
observe even-odd parity effects, see Fig.~\ref{fig5}.
For odd (even) $N$,
$G_1(V)$ exhibits a dip (peak) near zero bias. However, peaks
do not reach the unitary limit ($G_1=2e^2/h$) and
dips are not associated with perfect conductance zeroes anymore.
We call this behavior \textit{``weak parity effect''} to contrast it
from the ideal ``strong parity effect'' found for tunneling
into a Majorana chain or loop, see Sec.~\ref{sec3}.  
Notice that the even-odd features,
clearly visible in the lower inset in Fig.~\ref{fig5}
for the linear conductance, gradually disappear as $N$ increases,
as expected from the above argument.  The upper right
inset of Fig.~\ref{fig5} shows that the
linear conductance $G_1(0)$ monotonically increases
with the effective parameter $A(0)=\Omega\Gamma/t^2$.
The tunnel couplings in our model enter $G_1(0)$ in this
configuration only through $A(0)$.

For the \textit{closed ladder}, we find a very similar picture.
Now the solution to Eq.~(\ref{xxxx}) follows by Fourier expansion,
\begin{equation}\label{closedla}
X_1(\epsilon)=  \frac{1}{N}\sum_{m=-N}^{N-1}  \frac{1}{1-2iA(\epsilon)
\sin(\pi m/N)}.
\end{equation}
In the large-$N$ limit, this expression turns into an $N$-independent
integral and parity effects are quenched.  Numerical evaluation
[based on Eqs.~(\ref{diffcondzero}), (\ref{ga1a1}) and (\ref{closedla})]
shows that that the differential conductance $G_1(V)$ has $P$ peaks,
where the number $P$ coincides with the one found in
the homogeneous loop configuration, see Eq.~(\ref{Ploop}). However, now
the peak structure is less pronounced.

To conclude, a Majorana junction with ladder topology, either of closed
or open type, is found to exhibit a weak even-odd parity effect
that disappears for $N\to \infty$.

\subsection{Tree configuration}\label{sec4b}

Next we study the ``tree'' topology in Fig.~\ref{fig2}, where
$\gamma_{B,1}$ is connected to all other $\gamma_{B,j>1}$
through $\Omega_{1,j}=\Omega$, but the remaining $\hat\Omega$-couplings
vanish. Now Eq.~(\ref{xx}) has the solution
\begin{equation}
X_1(\epsilon)= \left( 1-\frac{(N-1)\Omega^2 (\epsilon+i\Gamma)^2}
{(\epsilon^2-t^2+i\epsilon\Gamma)^2}\right)^{-1}.
\end{equation}
The linear conductance then follows from Eq.~(\ref{lincond}),
\begin{equation}\label{treelin}
G_1(0)=\frac{2e^2}{h} \left(1+\frac{t^4}{(N-1)\Omega^2 \Gamma^2}\right)^{-1}.
\end{equation}
There is no oscillatory (even-odd) $N$-dependence,
and we find \textit{no parity effect} in the tree configuration.
Interestingly, as $N$ increases, $\gamma_{A,1}$ effectively
decouples and the unitary single-Majorana limit is approached.

\subsection{Isotropic configuration}\label{sec4c}

\begin{figure}[t]
\begin{center}
\includegraphics[width=0.9\textwidth]{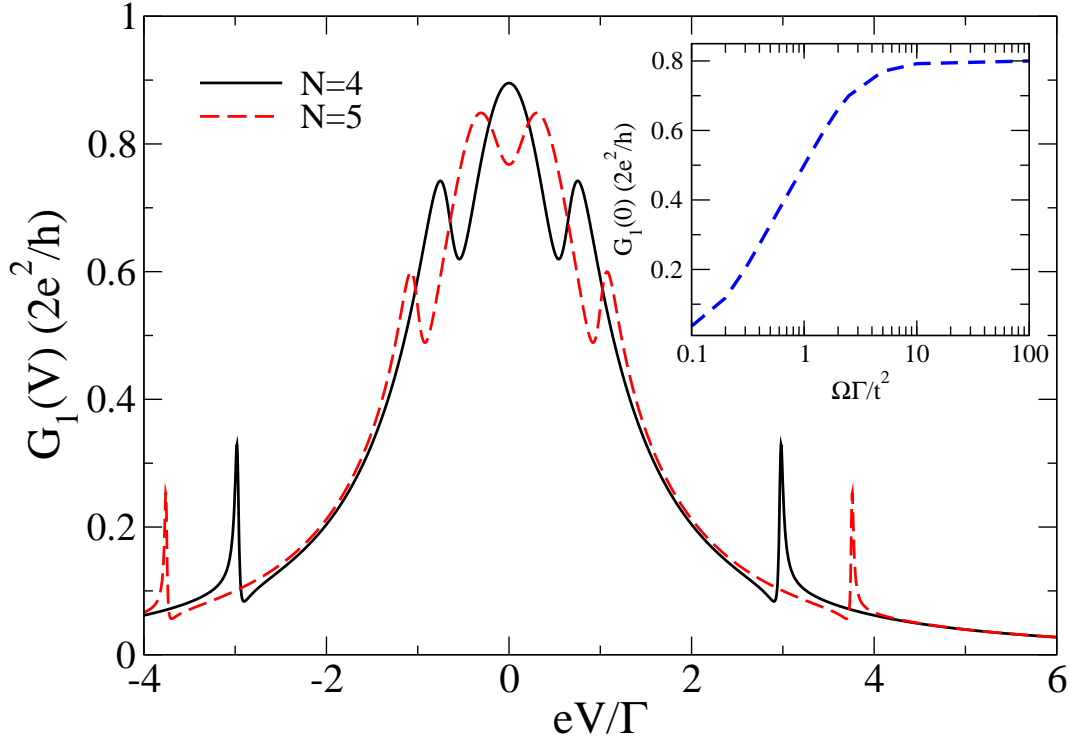}
\end{center}
\caption{\label{fig6}  $G_1(V)$ for the ``isotropic''
case with $N=4$ and $N=5$, for $t/\Gamma=0.5$ and $\Omega/\Gamma=1.2$.
Inset: Linear conductance $G_1(0)$ \textit{vs.}\ $A=\Omega\Gamma/t^2$
for $N=5$.
}
\end{figure}

Finally, consider the isotropic configuration in Fig.~\ref{fig2}, where
all couplings $\Omega_{j<k}=\Omega$ have the same magnitude.
(Note that similar isotropic couplings also arise through the charging interaction
\cite{beri2,alex}.)  As detailed in the Appendix,
for the isotropic Majorana junction with arbitrary $N$,
the tunneling conductance is given by
\begin{eqnarray}\nonumber
G_1(V) &=& \frac{2e^2}{h} {\rm Im} \left [ -\frac{\Gamma}{\epsilon+i\Gamma}
\left( 1+\frac{t^2 X_1(\epsilon)}
{\epsilon^2-t^2+ i\epsilon\Gamma} \right) \right]_{\epsilon=eV} ,
\\ \label{condiso}
X_1(\epsilon) &=& \frac{1}{1+ \left(1- \frac{1}{\xi_1(\epsilon)}\right )
A(\epsilon) } ,
\end{eqnarray}
where $A(\epsilon)= i\Omega(\epsilon+i\Gamma)/(\epsilon^2-t^2+i\epsilon\Gamma)$
is as in Eq.~(\ref{xxxx}) for the ladder topology, and
\begin{equation}\label{rec1}
\xi_1(\epsilon) = \frac12 \left[ 1+ \left(  \frac{1 +
 A(\epsilon)}{ 1 - A(\epsilon)} \right)^{N-1} \right].
\end{equation}
The nonlinear tunneling conductance $G_1(V)$ is shown in Fig.~\ref{fig6}.
It exhibits a weak parity effect, with a zero-bias peak  (dip) in
$G_1(V)$ for even (odd) $N$.  In particular, the dip does not lead to
a perfectly vanishing linear conductance.

To study this issue in more detail,
note that for $\epsilon=eV=0$ one has $A=\Omega \Gamma/t^2$
and the linear conductance is given by Eq.~(\ref{lincond}).
For $A\gg N$, Eq.~(\ref{rec1}) implies
$\xi_1\approx \delta_{N,{\rm odd}} + (-1)^{N-1} (N-1)/A$,
and hence we get in this limit $X_1 \simeq \delta_{N,{\rm odd}} / N$, i.e.,
\begin{equation}\label{extreme}
G_1(0) \simeq \frac{2e^2}{h}
\left ( 1 - \frac{\delta_{N,{\rm odd}}}{N} \right).
\end{equation}
The linear conductance thus exhibits a parity effect for
$\Omega\Gamma/t^2\gg 1$. However, this effect disappears for large $N$.
The approach to Eq.~(\ref{extreme}) with growing $A$ is
shown in the inset of Fig.~\ref{fig6} for $N=5$.

We conclude that the isotropic Majorana junction is characterized by a
\textit{weak parity effect} for small-to-intermediate $N$,
quite similar to the ladder configuration in Sec.~\ref{sec4a}.

\section{Discussion and conclusions}\label{sec5}

We now offer an intuitive and general interpretation of our results
for the linear tunneling conductance $G_1=G_1(0)$, where lead $j=1$
is connected to a set ${\cal M}$ of coupled Majorana fermions.
Every Majorana fermion in the set ${\cal M}$ is linked to other Majoranas in this set
through $K=1,2,\ldots$ ``bonds'', where a bond requires that the
respective tunnel coupling is finite.  In general, the set ${\cal M}$
contains not only the $\gamma_{\alpha,j}$ but also additional
Majorana fermions $\eta_j \propto c_j-c^\dagger_j$
corresponding to the normal parts (leads).  The $\eta_j$
are coupled only to $\gamma_{A,j}$ by $\lambda_j$, see
Eq.~(\ref{ham0}), with $\Gamma_j\propto \lambda_j^2$.
We note in passing that within the bosonization
approach, the $\eta_j$ represent
Klein factors for each lead \cite{beri2,alex}.
A Majorana mode with $K=1$ (but not $\eta_1$) is
called ``outer Majorana fermion'' and, if present, constitutes
a boundary of the set ${\cal M}$.
The tunneling conductance $G_1$ then follows by summing over all
propagation amplitudes for paths on ${\cal M}$
connecting $\eta_1$ and outer Majoranas, where a possible parity
effect follows from the following rule:
\textit{Trajectories containing an odd (even) number of
Majoranas in the set ${\cal M}$ yield a finite (zero) contribution to $G_1$.}
If ${\cal M}$ has no boundary (i.e., no outer Majoranas exist),
$G_1$ is determined by the sum over all ``loop'' trajectories
passing through $\gamma_{A,1}$.
In particular, ``homogeneous loop'' trajectories do not contribute to $G_1$,
see our analysis in Sec.~\ref{sec3b}.

We have verified this rule for the case of $N=3$ nanowires (for arbitrary parameters)
in Sec.~\ref{sec3a} as well as for all the examples in Sec.~\ref{sec4}.
For instance, in the ``tree'' configuration,
Eq.~(\ref{treelin}) shows that $G_1\ne 0$ for arbitrary $N$.
Since there are $N-1$ outer Majoranas $\eta_{j>1}$,
each connected to $\eta_1$ through a chain of four
Majoranas, every path connecting $\eta_1$ and $\eta_{j>1}$
involves an odd number (five) of Majoranas.
(Note that $\eta_1$ is not included in the set ${\cal M}$,
and hence, in the Majorana number, it should not be counted
while the outer Majorana $\eta_{j>1}$ counts.)
As a result,  a finite conductance $G_1$ follows from the above rule.
As another example, consider the ``ladder'' configuration with $N$ wires,
where we also have $N-1$ outer Majoranas $(\eta_{j>1})$.
Here, paths connecting $\eta_1$ and $\eta_m$, for
even $m=2,4,\ldots$, lead to an enhancement of $G_1$,
which can explain the reported peak \textit{vs.}\ dip structure for even
\textit{vs.}\ odd $N$.

The present work has shown that rich parity effects are present
in junctions of multiple Majorana nanowires.  For chain-like
arrangements, the $T=0$ linear tunneling conductance  is either zero
(for even number of Majoranas) or $2e^2/h$ (for odd number) \cite{flensberg}.
Such a strong parity effect does not survive for general junction
topologies.
For instance, in the ``ladder'' and ``isotropic'' topologies,
see Fig.~\ref{fig2}, one has only a ``weak parity effect'', where
non-ideal zero-bias dips and peaks can be observed in the tunneling
conductance depending on the parity of the Majorana number.
However, these effects are less pronounced and disappear with increasing number of Majoranas.
In other configurations, e.g., for the ``tree'' topology, there
is no parity effect at all.
A particularly noteworthy result concerns the tunneling conductance
when $\gamma_{A,1}$ is coupled to a ``Majorana loop.'' 
For generic parameters, a strong parity effect is present for such a
junction topology, with the even-odd features in the conductance
as for a Majorana chain.  However, for a homogeneous loop 
(and in some other special cases),
the linear conductance is found to vanish for any number
of Majorana fermions in the loop, i.e., no parity effect can be detected.

Finally, let us briefly address the role of finite-temperature effects
and of Coulomb interactions.  Finite $T$ implies a 
thermal broadening of all peak or dip features in the 
differential conductance, plus a reduction of the peaks.
Once $T$ reaches the energy scale corresponding to the separation
between two adjacent peaks in $G_1(V)$, peaks start to overlap and 
the parity effects described here will be smeared out.
Concerning Coulomb interactions, it has recently been pointed out
that charging effects cause interesting Kondo physics in similar
Majorana networks \cite{beri1,beri2,alex}.  This physics is different
from the parity effects studied here, and it would be interesting
to study their interplay with Kondo physics for finite charging energy.

\ack
We acknowledge support by the DFG within the research networks SFB TR 12
and SPP 1666, and by the Ministry of Science, Technology and Innovation
of Brazil.

\appendix
\section*{Appendix}

\subsection*{Retarded Majorana Green's function}

Here we briefly sketch the derivation of Eq.~(\ref{gret}), obtained
by solving the combined equations of
motion for Majoranas and lead fermions,
\[
i\partial_t \gamma_{A,j} = it_j \gamma_{B,j}+\lambda_j
\left (c_j^{}-c_j^\dagger  \right),   \quad
i\partial_t \gamma_{B,j} = -it_j \gamma_{A,j}+ i
\sum_{j'}\Omega_{jj'}\gamma_{B,j'},
\]
\[
(i\partial_t-\xi_k)c_{kj}= \lambda_j\gamma_{A,j}.
\]
The retarded Green's function is contained
in the general Keldysh Green's function
\[
{\cal G}_{\alpha j, \alpha' j'} (t,t') = -\frac{i}{\hbar} \left
\langle {\cal T}_C \gamma_{\alpha,j}(t) \gamma_{\alpha', j'}(t')\right\rangle,
\]
where ${\cal T}_C$ denotes time ordering along the standard Keldysh
contour \cite{alexa}.
Exploiting the wide-band approximation
for the leads, some algebra yields the Dyson equation for ${\cal G}$
from the above equations of motion,
\[
\left( \begin{array}{cc} \epsilon \hat 1 -\hat \Sigma(\epsilon) &
-i\hat t \\ i\hat t &\epsilon \hat 1- i \hat \Omega \end{array}\right)
{\cal G}(\epsilon) = \left( \begin{array}{cc}
\hat 1 & 0 \\ 0 & \hat 1 \end{array} \right).
\]
The $2\times 2$ structure refers to $\alpha=A,B$ space.
In addition, every matrix entry still carries the
standard $2\times 2$ Keldysh substructure \cite{alexa},
${\cal G} = \left( \begin{array}{cc} {\cal G}^r & {\cal G}^K \\
0 & {\cal G}^a \end{array} \right)$,
where ${\cal G}^a=[{\cal G}^r]^\dagger$ and the Keldysh component is
${\cal G}^K$.  The Keldysh structure of the self-energy
$\hat \Sigma(\epsilon)={\rm diag}(\Sigma_1,\ldots,\Sigma_N)$
due to the lead fermions is
\[
\Sigma_j (\epsilon) = - i \Gamma_j \left( \begin{array}{cc} 1 &
F(\epsilon - \mu_j) + F(\epsilon + \mu_j) \\
0 & -1 \end{array} \right) ,
\]
where $F(\epsilon)=\tanh(\epsilon/2k_BT)$.
Noting that $\hat\Omega$ and $\hat t$ are diagonal matrices in Keldysh space,
it is then straightforward to extract the $2N\times 2N$ matrix
for the retarded Green's function ${\cal G}^r$ quoted in Eq.~(\ref{gret}).

\subsection*{Isotropic junction}

Here we show how to derive the quantity $X_1(\epsilon)$ quoted
in Eq.~(\ref{condiso}) for the isotropic topology
in Sec.~\ref{sec4c}.  First, it is useful
to define auxiliary variables $Y_j(\epsilon) =\sum_{k=1}^j X_k(\epsilon)$
for $j=0,\ldots,N$, with $Y_0\equiv 0$. The
Dyson equation for $X_j$, see Eq.~(\ref{xx}), then takes the form
\[
(1+A)Y_j-(1-A)Y_{j-1}= \delta_{j,1} + A Y_N ,
\]
with $A(\epsilon)$ specified in Eq.~(\ref{xxxx}).
For $j=1$, this reduces to
\[
Y_1= \frac{1+AY_N}{1+A},
\]
while for $j>1$, we arrive at a non-homogeneous linear
recursion relation for the quantities $\xi_j\equiv Y_j/Y_N$,
\[
(1+A)\xi_j-(1-A)\xi_{j-1}=A, \quad \xi_N=1.
\]
One can easily verify that this set of equations is solved by
\[
\xi_j = \frac12 \left[ 1+ \left(  \frac{1 +
 A}{ 1 - A} \right)^{N-j} \right],
\]
where Eq.~(\ref{rec1}) follows for $j=1$.  Now note that $X_1=Y_1=\xi_1 Y_N$
but also $Y_1=(1+AY_N)/(1+A)$, see above.  We  thereby obtain
$X_1(\epsilon)$ in the quoted form [Eq.~(\ref{condiso})], where
$G_1(V)$ follows from Eqs.~(\ref{diffcondzero}) and (\ref{ga1a1}).

\end{document}